# ATLAS MDT TDC Simulations for LHC Run3 and HL-LHC

J. Ge[a], B. Li[b], M. Fras[c], J. Zhu[a], B. Zhou[a], T. Dai[a,*]

[a] *Department of Physics, University of Michigan, Ann Arbor, MI, 48109, USA*
[b] *Institute of Frontier and Interdisciplinary Science and Key Laboratory of Particle Physics and Particle Irradiation (MOE), Shandong University, Qingdao, China*
[c] *Max-Planck-Institut für Physik, Werner Heisenberg Institute, Boltzmannstr. 8- 85748 Garching*

 \* *Corresponding author. E-mail*: `daits@umich.edu`

ABSTRACT: The Large Hadron Collider (LHC) started the Run 3 operation in 2022, and the peak instantaneous luminosity in Run 3 may reach $3\times10^{34}$ $cm^{-2}s^{-1}$. The ATLAS Monitored Drift Tube (MDT) chambers are the main component of the precision tracking system in the ATLAS muon spectrometer. It is important to understand any potential issues with the MDT Front-End (FE) readout electronics for an expected level-1 (L1) trigger rate of 100 kHz and a complex deadtime of over 5% for Run 3 operations. We use raw data collected in 2022 to emulate the expected hit rates in MDT chambers and perform a realistic simulation on the ATLAS Muon TDC (Time-to-Digital Converter) (AMT) chip with the current configuration. We study the AMT chip performances by analyzing the trigger/L1/readout buffer occupancies and hit loss fractions under different luminosities with L1 rate of 100 kHz by using the Modelsim software. The hit loss fraction of the hottest MDT chamber (BIL3C05) is lower than 5% due to FE readout, even at a luminosity of $5.01\times10^{34}$ $cm^{-2}s^{-1}$ with a deadtime of 5% and a L1 rate of 100 kHz, indicating that AMT can operate under Run 3 conditions without problems. The MDT trigger and readout electronics will be replaced for triggerless readout during High-Luminosity LHC (HL-LHC) runs. We also simulate the AMT behavior in the triggerless mode up to $7.44\times10^{34}$ $cm^{-2}s^{-1}$ and propose possible AMT configurations in case some FE electronics could not be replaced during the long shutdown 3 (LS3).

# Contents

## 1   Introduction

Monitored Drift Tubes (MDT) plays an important role in the ATLAS muon spectrometer to detect charged particles from proton–proton collisions. The MDT chambers are engineered to provide high-precision tracking of charged particles within the pseudorapidity region |η| < 2.7, achieving average spatial resolutions of 80 µm per tube and 35 µm per chamber. They can operate at hit rates of up to 300 kHz per tube without any significant degradation in either spatial resolution or detection efficiency. [1-3]. Each MDT chamber has up to 18 Front-End Electronics (FEE) boards ("mezzanine cards"), which contains three custom-designed 8-channel Amplifier/Shaper/Discriminator (ASD) chips and one 24-channel Time-to-Digital Converter (TDC) chip (called AMT here, the current version is AMT-3) [4-5]. An on-chamber data-acquisition (DAQ) board called Chamber Service Module (CSM) reads out the data of all mezzanine cards on a chamber. The CSM packages and serializes the data and then sends these data to a far-end MDT Readout Driver (MROD) via an optical fiber [6-7].

So far, the ATLAS MDT chambers as well as their electronics have been running smoothly for nearly 15 years. In 2022, the Large Hadron Collider (LHC) started Run 3 after Long Shutdown (LS) 2 and a instantaneous luminosity above $2\times10^{34}$ cm$^{-2}$s$^{-1}$ has been achieved [8]. The current ATLAS MDT electronics were designed to operate under the LHC luminosity of $1\times10^{34}$ cm$^{-2}$s$^{-1}$ but is still being used in Run 3 [1]. The accepted Level 1 (L1) trigger rate will reach 100 kHz in Run 3, and the deadtime increases exponentially with L1 rate, exceeding 4% [9]. In 2030, the ATLAS detector will take data in the High Luminosity LHC (HL-LHC) period after Phase-II upgrade. The instantaneous luminosity may ultimately be pushed up to $7\sim7.5\times10^{34}$ cm$^{-2}$s$^{-1}$ during HL-LHC period [10]. The targeted ATLAS accept rate and latency at L1 level for HL-LHC are 1 MHz and 10 µs, respectively [11]. In such a high trigger rate, the new MDT TDCs will work in triggerless mode to avoid sending same hit for repeatedly, while the current MDT TDCs work in triggered mode and sends out all hits inside a 1.3 µs time window upon arrival of an L1 Trigger Accept (L1A) signal [11-12]. It's expected to replace all 15164 MDT legacy mezzanines with newly developed ones for HL-LHC runs during LS3. However, it is possible that the MDT mezzanines on some chambers could not be replaced due to limited accessibility, albeit in a very small fraction of cases [13]. It means that some current AMT-3 will likely continue to work in triggerless mode during the HL-LHC period.

We need to verify if the AMT-3 will have any issues at high luminosities with 100 kHz L1A rate in Run 3, and if the AMT-3 on some MDT chambers can meet the demand in HL-LHC runs. In this paper, we conducted the verification of the AMT-3 performance at high luminosities for Run 3 and HL-LHC runs. We processed the 2022 raw data of ATLAS MDT to get the hit input of AMT-3 at high luminosity. And then we did AMT-3 behavioral level simulation using its Verilog chip code with Modelsim software [14]. We will present our simulation method along with the results.

## 2   AMT-3

The AMT-3 chip can measure the arrival time of leading- and trailing-edge of the output hits from 24 channels of three ASD chips. Since there is a Wilkinson integrator in each channel of the ASD chip, the energy of the pulse can also be acquired by measuring its width (time of trailing-edge subtracts time of leading-edge). The architecture of the AMT-3 chip is shown in Figure 1 [4, 6, 15]. The clock for AMT-3 is the 40 MHz LHC clock, and a Phase Locked Loop (PLL) circuit multiplies the frequency to 80 MHz for coarse time counter. 16 timing signal taps (0.78 ns per tap) from a voltage-controlled oscillator (Asymmetric Ring Oscillator) are used for fine-time measurement. The average time measurement resolution of the 24 channels is about 250 ps [6].

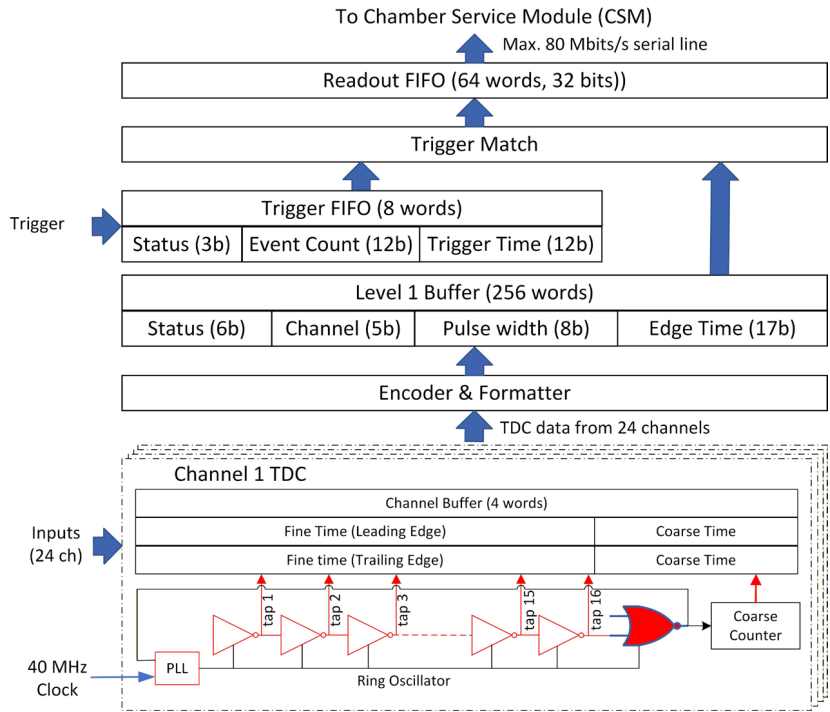

**Figure 1.** Block diagram of the AMT-3 chip [5].

Time measurement results of each channel are stored in a channel buffer (4 words deep). Then the leading and trailing edges of each channel are paired and saved to a common Level 1 (L1) buffer (256 words deep). In triggered mode, the L1 trigger is input to the AMT-3 via an eight-words trigger First In-First Out memory (FIFO). The AMT-3 will search for hits stored in the L1 buffer within a programmable time-window beginning at a given trigger, which is done via a matching function. The matched data are stored in a readout FIFO (64 words deep) and then transferred to the CSM serially in speed 10 ~ 80 Mbps. Although the AMT-3 is designed for operation at the present luminosities and in triggered mode, we investigate its possible use at higher luminosities in a triggerless mode. In triggerless mode, the trigger matching is disabled and all data from the L1 buffer is passed directly to the read-out FIFO. Hence the FIFO size of the AMT-3 is 320 words in triggerless mode [5].

The AMT-3 configuration and debugging are done by the CSM via the JTAG interface. The ASD chip setup is also handled by the AMT-3. Under the current configurations, AMT-3 is working in triggered mode, and the time window is set as 1300 ns. Both leading- and trailing- edge times of a hit are measured and transmitted. The serial output transmission speed is 80 Mbps.

## 3 Simulation method

The workflow of the AMT-3 simulation is shown in Figure 2. To simulate the input for the AMT-3 chip, we generated the corresponding input files from MDT raw data recorded during Run 3 in 2022. The data were collected in triggered mode during 13.6 TeV collisions, with a time window of 1300 ns, and contain both hit and trigger information per chamber. The information of each ATLAS run, and the luminosities of each Luminosity Block (LB) in a run can be found on "ATLAS Data Summary 2022-pp" web page [16]. We select multiple ATLAS runs with different peak stable luminosities for generating simulation input files. The runs have the same collision bunch number of 2450. Then a period of continuous LBs that have similar luminosities in an ATLAS run was intercepted from the whole run to get the hits at a specific luminosity (average value of the luminosities of selected LBs). The luminosities from the raw data of ATLAS runs range from $1.08 \times 10^{34}$ cm$^{-2}$s$^{-1}$ to $2.43 \times 10^{34}$ cm$^{-2}$s$^{-1}$ (maximum LHC luminosity in 2022 runs).

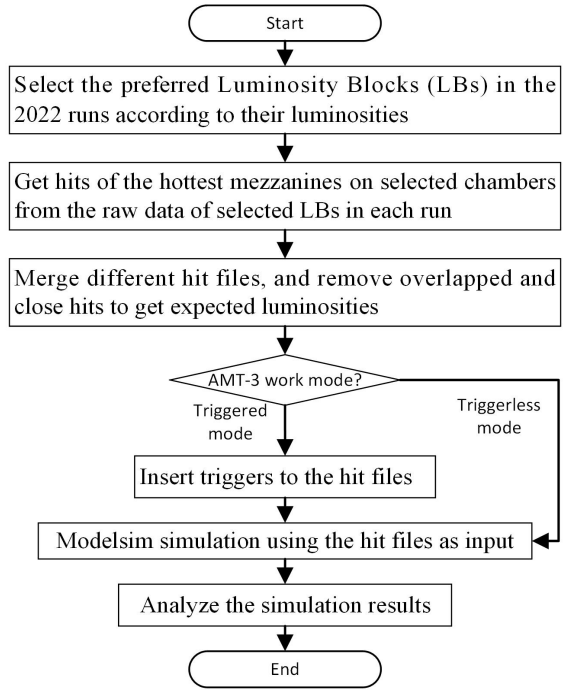

**Figure 2.** Workflow of the AMT-3 simulation.

To simulate the AMT-3 performance in Run 3, we selected the 2022 data of two chambers, one with the highest hit rate in all chambers (BIL3C05), and the other has the highest hit rate in the end-cap chambers (EML1A05). To assess the feasibility of operating the AMT-3 in triggerless mode during the HL-LHC period, two chambers with significantly different expected hit rates were selected to encompass a wider range of operational conditions: one with a low rate from the BO region (BOL3A13) and one with a relatively high rate from the EM region (EMS4C04). Then we select the hottest mezzanine on each chamber. From the MDT raw data of selected LBs in each run, we can extract the leading and trailing times of the hits measured by the AMT-3 chips on the chosen mezzanines. We normalize the start time to 0 ns and save the time of all hits of the selected mezzanines in different runs as a list file. The definitions of each column of the list files are shown in Table 1. Then we get the hit files of the selected mezzanines and chambers with different luminosities. The duration of the hit files for Run 3 and HL-LHC simulation are 800 ms and 400 ms respectively.

**Table 1.** The definitions of each column of the hit files

| Column Number | 1 | 2 | 3 | 4 |
|---|---|---|---|---|
| Definitions | Channel Number of the AMT-3 Chip (0-23). | Arrival time of hit / trigger. Unit: ns. | 0: trailing edge of hit; 1: leading edge of hit. | 0: A hit; 1: A trigger |

To generate the hit files with luminosities higher than $2.43 \times 10^{34}$ $cm^{-2}s^{-1}$, we merge multiple hit files since the average hit rate per channel of the same mezzanine is proportional to the luminosity. The luminosity of the merged hit files is basically the sum of the luminosities of the original hit files. The luminosities of the hit files for Run 3 simulations are in the range from $1.08 \times 10^{34}$ $cm^{-2}s^{-1}$ to $5.01 \times 10^{34}$ $cm^{-2}s^{-1}$ while the luminosities of the hit files for HL-LHC simulations are in the range from $2.98 \times 10^{34}$ $cm^{-2}s^{-1}$ to $7.44 \times 10^{34}$ $cm^{-2}s^{-1}$. The step of the luminosities is about $0.5 \times 10^{34}$ $cm^{-2}s^{-1}$. Then we remove the overlapped hits, e.g. the series of the same channel like: leading edge, leading edge, trailing edge, trailing edge. And we also remove the hit that is too close to the previous one in the same channel (interval less than 500 ns). The average hit rates per channel of the selected mezzanines on the four chambers with different luminosities are shown in Figure 3. The average hit rates of BIL3C05 and EML1A05 at luminosity of $5.01 \times 10^{34}$ $cm^{-2}s^{-1}$ are about 220 kHz and 135 kHz, respectively. The average hit rates of BOL3A13 and EMS4C04 at luminosity of $7.44 \times 10^{34}$ $cm^{-2}s^{-1}$ are about 35 kHz and 80 kHz, respectively.

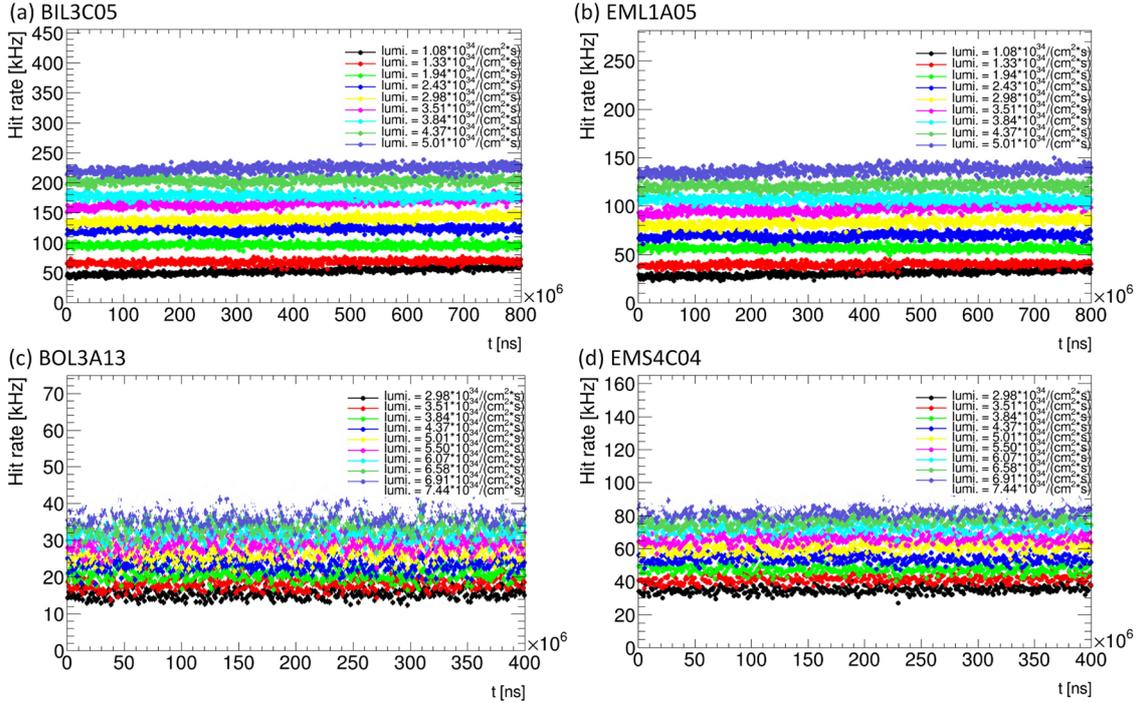

**Figure 3.** The hit rates per channel of the selected mezzanines on BIL3C05 (a), EML1A05 (b), BOL3A13 (c), and EMS4C04 (d) with different luminosities.

For the HL-LHC conditions, we can use the current hit files as the input of the simulation since the TDC will work in triggerless mode. However, for the Run 3 simulations, triggers must be inserted to the hit files as a part of the input. To prevent the front-end buffers from overflowing, the Central Trigger Processor (CTP) in ATLAS produces two type of deadtimes: simple deadtime and complex deadtime [17]. The simple dead-time is a fixed number of Bunch Crossing (BC) after each L1A (the current configuration is 4 BCs, equal to 100 ns). The complex dead-time mechanism is represented using a leaky-bucket model to simulate the behavior of a front-end buffer. In this framework, dead-time occurs whenever the bucket reaches its maximum capacity. The parameter S (expressed in units of L1A) denotes the effective depth of the bucket, corresponding to the available buffer size in the front-end electronics. The parameter R (measured in BC) specifies the leakage rate, i.e., the interval required for the bucket to drain the equivalent of one L1A. When the bucket if full, the new L1As will be vetoed, until the deadtime (R BCs) is end. The current ATLAS deadtime configuration features four distinct complex deadtime settings: bucket 0: 15/370, bucket 1: 42/384, bucket 2: 7/351, and bucket 3: 14/260. Among these, bucket 2 (7/351) is the most stringent deadtime setting and it constitutes the predominant source of the overall system deadtime in actual operation [18]. An example of the complex deadtime in real case scenerios (S/R = 7/351) is shown in Figure 4 (a). The average trigger period is limited to R BCs. The L1A rate is about 100 kHz in Run 3.

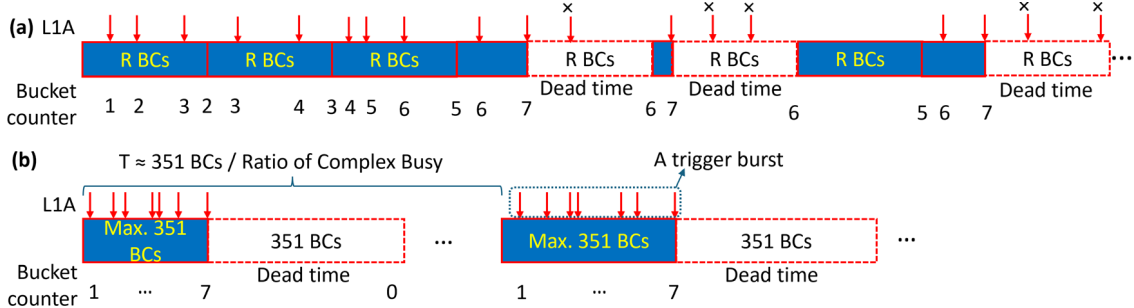

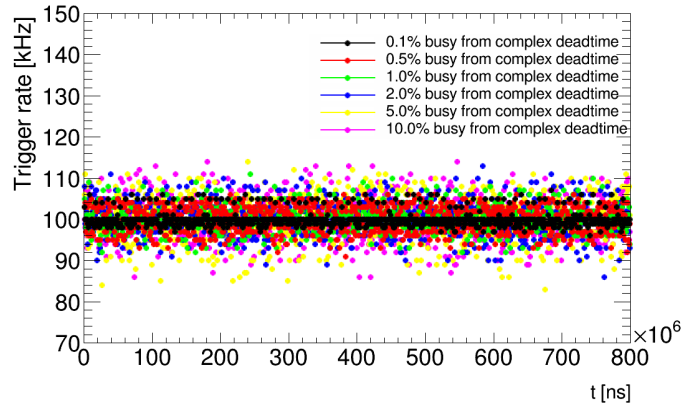

**Figure 4.** Examples of the complex deadtime in real case (a) and in extreme case for simulation (S/R = 7/351), T refers to the average time interval (Poisson distribution) between two adjacent trigger blocks (b).

**Figure 5.** Trigger rates at different ratios of complex busy.

In the case of our simulation, we generate Poisson-distributed triggers with average 100-kHz frequency and 100-ns simple deadtime after each trigger at first. We then generate triggers incorporating the complex deadtime using a simplified leaky bucket model, implemented in two steps: (1) Generate triggers for the extreme case: one block with 7 Poisson-distributed L1As within 351 BCs, followed by a 351-BC complex deadtime period; (2) Integrate blocks into the trigger stream: Insert these blocks into the nominal 100 kHz trigger stream following a Poisson distribution. The insertion probability for each block is set equal to its corresponding complex busy ratio (i.e., the L1 busy fraction due to complex deadtime). The evaluated ratios are: 0.1%, 0.5% (approximate average case), 1.0%, 2.0%, 5.0% (extreme case), and 10% (a possible instantaneous peak). An example of the complex deadtime in simulation is shown in Figure 4 (b). Finally the simple triggers are randomly deleted to rescale total trigger rate to 100 kHz. The trigger rates of different ratios of complex busy are shown in Figure 5. It can be observed that both the average trigger rate and the frequency of trigger bursts (A group of 7 L1As within 351 BCs) increase with the ratio of complex busy. These bursts consequently exert higher pressure on the AMT-3 buffers. The AMT-3 will have no problem in ATLAS if the AMT-3 works in above extreme case.

After generating the hit files with different combinations of chambers, luminosities and the ratios of complex busy, we can start the simulation. Since the AMT-3 is a digital chip designed by Verilog Hardware Description Language (HDL), we use the HDL simulator to verify the performance of the chip under different conditions. Modelsim is a verification and simulation tool for multiple languages, and it can simulate at behavioral, register-transfer level (RTL), or gate level [14]. We use Modelsim to conduct behavioral simulation of AMT-3. The simulation library is exactly the same with the one of the AMT-3 fabrication (Toshiba TC220G) [5]. Therefore, the simulation model can be as similar as possible to the real AMT-3 chip.

A Modelsim testbench was created to configure the AMT-3 simulation model, set up the input signal, and readout the output of the simulation. The configuration of the AMT-3 simulation model is exactly the same with the real situation. The serial output speed is 80 Mbps. In the triggered mode, the time window is set as 1300 ns. In both triggered and triggerless mode simulation, the AMT-3 model conducts combined measurement of leading and trailing edge. In trigerless mode, we also simulate the case of the single leading-edge measurement. The single measurement will generate half the data compared to the combined measurement [5]. Therefore, single measurement mode can reduce buffer occupancy to reduce hit loss. But the single measurement will lose the energy data of the hits. However, if the AMT-3 will still remain on some chambers, they will probably have to work in single measurement mode.

The testbench then generate the pulse according to the time of leading- and trailing- edge in the hit files and input them to the corresponding channel. At the arrival of the hits, the AMT-3 model will start time measurement and data record processes. When operating in triggered mode, the test bench supplies the trigger signal to the trigger port according to the timestamps defined in the input files. Upon trigger arrival, the corresponding BC is written into the trigger FIFO, and the AMT-3 initiates trigger matching. All data in the trigger matching window are written into the read-out FIFO and will be transmitted serially from the output port. At last, the testbench parallels the output data, and the measured time of the leading and trailing edges of the hits will be extracted and saved to files. The occupancies of the L1 buffer, trigger FIFO, and readout FIFO will also be read out by the testbench and saved for analysis.

## 4 Simulation results

### 2.1 AMT-3 simulation for Run 3 (triggered mode)

We simulated the AMT-3 performance on chamber BIL3C05 and EML1A05 at different luminosities and ratios of complex busy. The AMT-3 worked on triggered mode. We analyzed the buffer occupancies and the hit loss fractions from the output files of simulation. The hit loss can be calculated by:

$$Hit\ loss\ fraction = 1 - \frac{Number\ of\ measured\ hits}{Number\ of\ expected\ hits} \qquad (4.1)$$

The numbers of measured hits can be obtained from the output files of the simulations. In triggered mode simulation, the hits within 1300-ns time window after a trigger in each hit file are counted as the expected hits, while in triggerless mode simulation, all the hits are counted.

We take the case when the ratio of complex busy is 5% as an example and show the buffer occupancies at different luminosities in Figure 6 (a-c). Actually, for the same chamber, the higher luminosity and more burst triggers, the higher the L1 buffer, trigger FIFO, and readout FIFO occupancies. There are only readout FIFO overflow happened even in the hottest chamber at a complex busy ratio of 10% and luminosity of 5.01 x$10^{34}$ cm$^{-2}$s$^{-1}$. The hit loss is strongly correlated to the buffer overflows. A higher buffer overflow will result in a higher hit loss. BIL3C05 is the hottest chamber now. The hit loss fraction is lower than 5% when the luminosity reaches 5.01x$10^{34}$ cm$^{-2}$s$^{-1}$ and the ratio of complex busy is 5%, which is perfectly acceptable for our MDT operation. The hit loss fractions of the EML1A05 chamber are lower than 0.1% in general even at luminosity of 5.01x$10^{34}$ cm$^{-2}$s$^{-1}$ and the complex busy ratio of 10%, which are tiny and can be ignored overall. The results show that the AMT-3 can run without issues in Run 3.

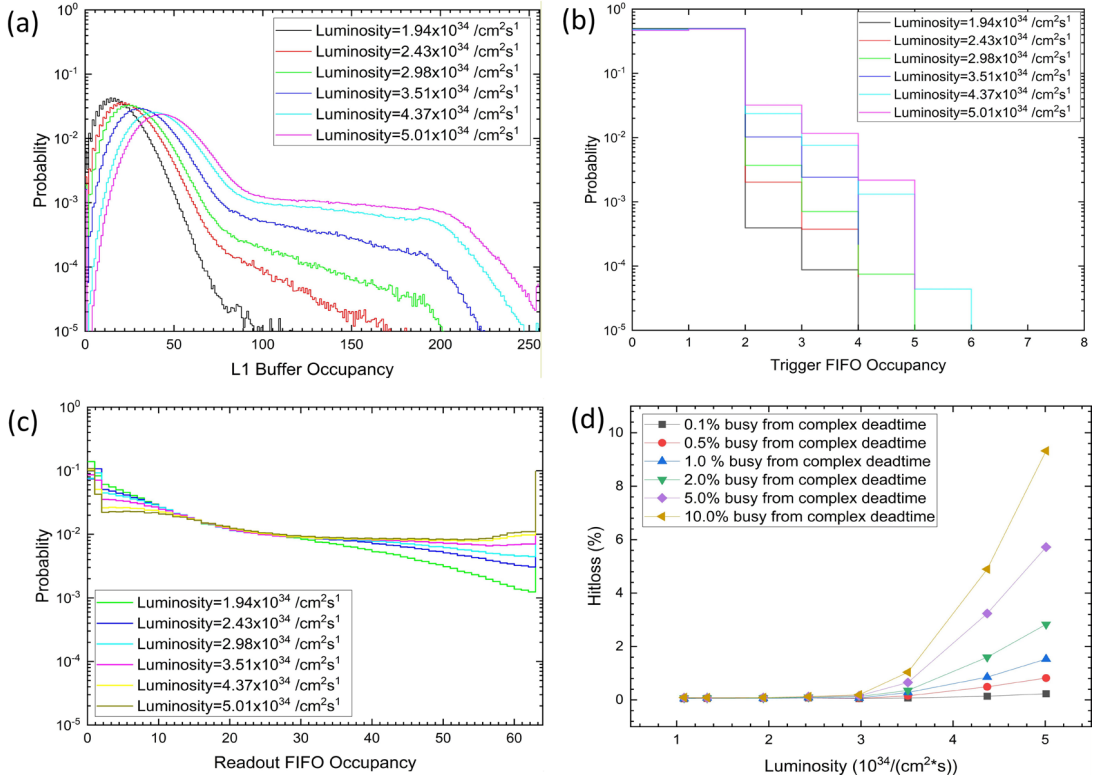

**Figure 6.** Chamber: BIL3C05, Probability distribution for: L1 buffer occupancies (a), trigger FIFO occupancies (b), and readout FIFO occupancies (c) of 5% complex busy at different luminosities, and hit loss fractions at different ratios of complex busy and luminosities (d).

## 2.2 AMT-3 simulation for HL-LHC (triggerless mode)

As for AMT-3 simulation for HL-LHC, we also analyzed the hit loss fractions at different luminosities. In triggerless mode, we must focus on hit latencies, which are latencies between the hit arriving at AMT-3 and the measuring data output. In ATLAS Run 4, MDT hits should be received by Level-0 MDT trigger (L0MDT) within 2.8 µs to be used by L0MDT. And only the hits received in the 10 µs L0 latency can be readout [10-11]. By considering latencies contributed by the time of flight (maximum about 140 ns), the drift time in the tube (maximum about 700 ns), and the fiber from CSM to L0MDT (maximum about 110 m, with latency about 550 ns) [1, 6, 10]. The hits should be sent out by the MDT TDC within roughly 1.4 µs and 8.6 µs to be used by the L0MDT and for readout, respectively.

Figure 7 and 8 show the hit loss fractions, hit latency distributions, and fractions of hit delay over 1.4 µs and 8.6 µs at different luminosities of BOL3A13 and EMS4C04 both on edge mode (with leading and trailing edge both measured) and leading-edge mode. Considering the results shown in Figure 3 (c) (d), Figure 7 and 8 (a) (d), we can observe that the hit loss fraction is ignorable under the hit rate at around 40 kHz of edge mode and at around 80 kHz of leading-edge mode, and will increase rapidly when the hit rate exceeds these critical points, which is consistent with our theoretical expectation. So the AMT-3 on BOL3A13 can meet the requirement of hit loss at luminosity of $7.44 \times 10^{34}$ $cm^{-2}s^{-1}$ when working in triggerless edge mode and leading-edge mode, while the AMT-3 on EMS4C04 can only work in leading-edge mode to fulfil the requirement of hit loss. The conclusion can be drawn from Figure 3 (c-d), Figure 7 and 8.

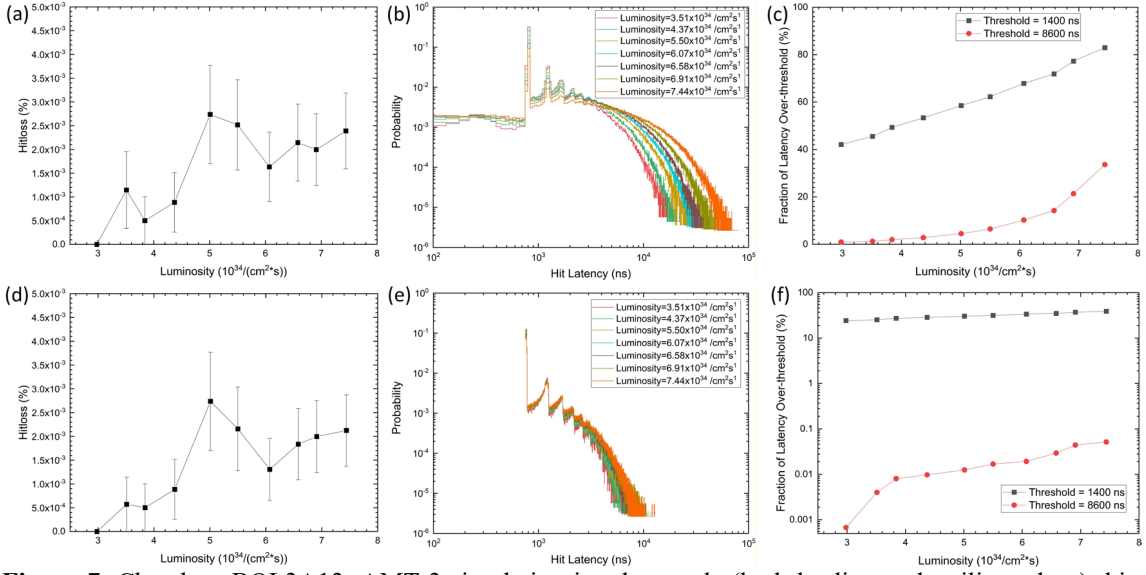

**Figure 7.** Chamber: BOL3A13. AMT-3 simulation in edge mode (both leading and trailing edges): hit-loss fraction (a), hit-latency distributions (b), and fractions of hit latency above 1.4 μs and 8.6 μs (c) at different luminosities.
AMT-3 simulation in leading-edge mode: hit-loss fraction (d), hit-latency distributions (e), and fractions of hit latency above 1.4 μs and 8.6 μs (f) at different luminosities.

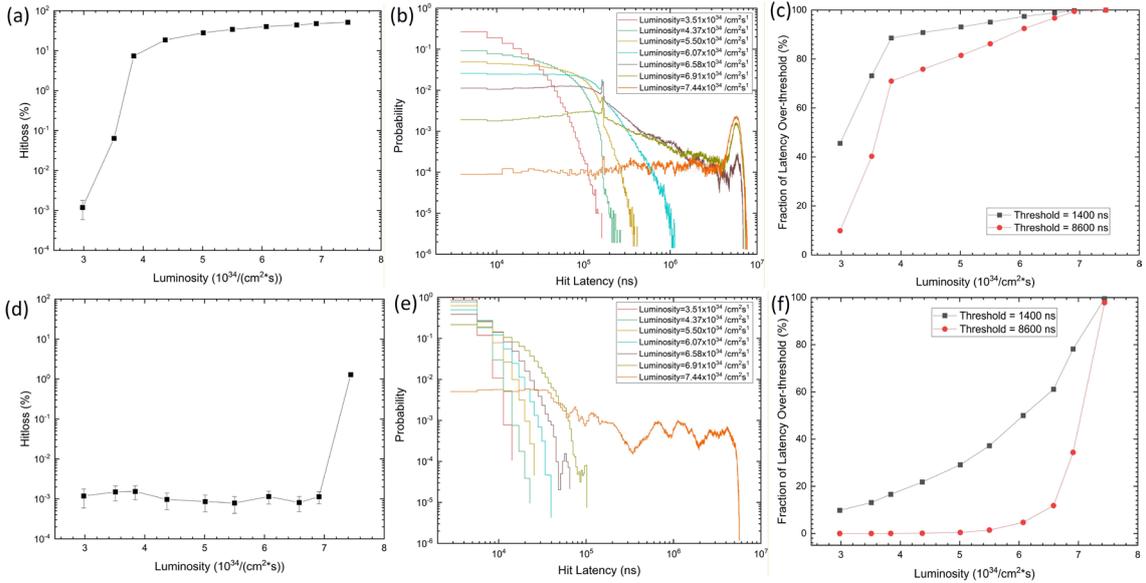

**Figure 8.** Chamber: EMS4C04. AMT-3 simulation in edge mode (both leading and trailing edges): hit-loss fraction (a), hit-latency distributions (b), and fractions of hit latency above 1.4 μs and 8.6 μs (c) at different luminosities.
AMT-3 simulation in leading-edge mode: hit-loss fraction (d), hit-latency distributions (e), and fractions of hit latency above 1.4 μs and 8.6 μs (f) at different luminosities.

However, based on the hit-latency distributions shown in Figures 7 and 8 (c) and (f), the proportion of hits with latencies exceeding 1.4 μs can approach 30% at high luminosities, even when the AMT-3 operates in leading-edge mode under very low hit-rate conditions (e.g., BOL3A13). So the chambers with the old mezzanines basiclly can not be used as L0MDT. The fractions of hit latency over 8.6 μs of EMS4C04 and BOL3A13 are about 30 % and 100% respectively when working on edge mode. It shows that the hits from the chambers with old mezzanines can not even be readout when AMT-3 is working on edge mode. From our simulation in Figure 7 (f), AMT-3 can fulfil the 10-μs L0 latency requirements in HL-LHC period under leading-edge mode when the hit rate is low. That means the hits from the low hit-rate chambers with AMT-3 can be readout.

## 5  Conclusions

The simulations to verify AMT-3 performance in Run 3 and HL-LHC periods were done. We used raw data collected in 2022 to emulate the expected hit rates in MDT chambers. By combining the hit files, we can get the hit files with luminosities higher than the current runs. For triggered mode simulation, we inserted 100 kHz trigger with the ratios of complex busy from 0.1% to 10% to the hit files. Using the hit files as the input, we conducted Modelsim behavioral simulation of AMT-3 with the current AMT-3 configuration in ATLAS. The simulation model of AMT-3 is constructed by AMT-3 verilog code and its original libraries.

For the AMT-3 triggered mode simulation for Run 3, we analyzed the trigger/L1/readout buffer occupancies and hit loss fractions under different luminosities with L1 rate of 100 kHz at different ratios of complex busy. We found that higher luminosities and more burst triggers will cause higher buffer occupancies. Then the buffer overflow will result in hit loss. The hit loss fraction of the hottest MDT chamber (BIL3C05) is lower than 5% even at a luminosity of 5.01 x$10^{34}$ cm$^{-2}$s$^{-1}$ with a complex busy ratio of 5% and a L1 rate of 100 kHz, indicating that AMT can operate under Run 3 conditions without problems. We also simulate the AMT behavior in the triggerless mode with luminosities up to 7.44x$10^{34}$ cm$^{-2}$s$^{-1}$ for HL-LHC period in case some FE electronics could not be replaced during the long shutdown 3 (LS3). From the perspective of the hit latencies, the hits measured by AMT-3 can not be used as L0MDT. However, AMT-3 can fulfil the 10-μs L0 latency requirements in HL-LHC period under leading-edge mode when the hit rate is low. That means the hits from the low hit-rate chambers with AMT-3 can be readout within the L0 latency in HL-LHC runs.

### Acknowledgement

We warmly acknowledge all our ATLAS and CERN colleagues without whom the ATLAS muon system could not have been built. Furthermore, we are grateful to the funding from DOE (DE-SC0012704, DE-SC0007859), United States of America which supported our MDT Maintenance and Operation work.